\documentclass[12pt]{article}

\def\beq{\begin{eqnarray}}
\def\eeq{\end{eqnarray}}

\def\lsim{\mathrel{\rlap{\lower3pt\hbox{\hskip0pt$\sim$}}
     \raise1pt\hbox{$<$}}}         
\def\gsim{\mathrel{\rlap{\lower4pt\hbox{\hskip1pt$\sim$}}
     \raise1pt\hbox{$>$}}}         

\usepackage{amsmath}
\usepackage{amsfonts}

\begin{document}
\begin{titlepage}


\medskip
\medskip

\centerline{\Large \bf Non-perturbative Unitarity of Gravitational Higgs Mechanism}
\medskip

\centerline{\large Alberto Iglesias$^\dag$\footnote{\tt Email: ai372@nyu.edu}
and Zurab Kakushadze$^\S$\footnote{\tt Email: zura@quantigic.com}}

\bigskip

\centerline{\em $^\dag$ Center for Cosmology and Particle Physics,
New York University}
\centerline{\em 4 Washington Place, New York, NY 10003}
\medskip
\centerline{\em $^\S$ Quantigic$^\circledR$ Solutions LLC}
\centerline{\em 200 Rector Place, 43C, New York, NY 10280
\footnote{DISCLAIMER: This address is used by the corresponding author for no
purpose other than to indicate his professional affiliation as is customary in
scientific publications. In particular, the contents of this paper are limited
to Theoretical Physics, have no commercial or other such value,
are not intended as an investment, legal, tax or any other such advice,
and in no way represent views of Quantigic$^\circledR$ Solutions LLC,
the website
{\underline{www.quantigic.com}}
or any of their other affiliates.}}
\medskip
\centerline{(February 21, 2011)}

\bigskip
\medskip

\begin{abstract}
{}In this paper we discuss massive gravity in Minkowski space via
gravitational Higgs mechanism, which provides a non-perturbative definition thereof.
Using this non-perturbative definition, we address the issue of unitarity
by studying the full nonlinear Hamiltonian for the relevant metric degrees of freedom.
While perturbatively unitarity is not evident, we argue that no
negative norm state is present in the full nonlinear theory.

\end{abstract}
\end{titlepage}

\newpage

\section{Introduction and Summary}

{}The gravitational Higgs mechanism gives a non-perturbative and fully
covariant definition of massive gravity \cite{KL, thooft, ZK, Oda, ZK1, ZK2,
Oda1, Demir, Cham1, Oda2, JK, IgKa, Cham2, Ber2, Cham3}. The graviton acquires mass via spontaneous
breaking of the underlying general coordinate reparametrization invariance by
scalar vacuum expectation values.\footnote{For earlier and subsequent related
works, see, {\em e.g.}, \cite{Duff, OP, GMZ, Perc, GT, Siegel, Por, AGS,
Ch, Ban, AH1, CNPT, AH2, Lec, Kir, Kiritsis, Ber, Tin, Jackiw, Tin1, RG, HHS, RG1},
and references therein.} In this paper, motivated by the results of
\cite{IgKa} for the gravitational Higgs mechanism in de Sitter space, we
study the issue of unitarity of the gravitational Higgs mechanism in the Minkowski background within the setup of \cite{ZK1}.

{}Within the {\em perturbative} framework, unitarity requires that, in order not to propagate a negative norm
state at the quadratic level in the action, the graviton mass
term be of the Fierz-Pauli form \cite{FP}. Furthermore, higher order terms should be such that they do not introduce
additional degrees of freedom that would destabilize the background, and difficulties in achieving this to all orders
in perturbative expansion have been known for quite some time (see, {\it e.g.}, \cite{BD}, \cite{CNPT}, and references therein).
In this regard, in order to circumvent the aforesaid difficulties within the perturbative expansion, \cite{RG} proposed an order-by-order construction, albeit in a special decoupling limit, such that higher-than-second-order time derivatives in the equations of motion are absent.

{}Here we propose a different approach. Following \cite{IgKa}, our key observation is that
perturbation theory appears to be inadequate, among other things, for the purposes of
addressing the issue of unitarity.\footnote{For instance, in the de Sitter case, this becomes evident from the fact that in the
full nonlinear theory no enhanced local gauge symmetry (or ghost) is present for any value of the Hubble parameter, while its appearance
in the perturbative framework already at the quadratic order appears to be a mere artifact of linearization \cite{IgKa}.}
Thus, while the theories of \cite{ZK1} reproduce the Fierz-Pauli action at the quadratic level, according to \cite{Ber2}, at higher orders in perturbative expansion they do
{\em not} reduce in the decoupling limit to the theories studied in \cite{RG} (and, according to \cite{Ber2}, the same holds for the models discussed in \cite{Cham1}).
However, since the definition of \cite{ZK1} is intrinsically non-perturbative,
we can test the stability of the Minkowski background in the full nonlinear theory. In this paper we perform a non-perturbative Hamiltonian analysis for the relevant metric
degrees of freedom and argue that the full nonlinear theory appears to be free of
ghosts.

{}Our main result is that, in the gravitational Higgs mechanism, {\em non-perturbatively}, the Hamiltonian appears to be bounded from below in the Minkowski
background.\footnote{We arrived at the same conclusion in \cite{IgKa} in the de Sitter case.} We
argue that this is indeed the case\footnote{Here we should note that in this regard we only analyze in detail the simplest example with higher (namely, four) derivative couplings in the scalar sector (\ref{quad.pot}); however, based on the fact that in the gravitational Higgs mechanism diffeomorphisms are broken spontaneously, we believe our conclusions should hold in the general case as well, albeit we do not have a proof of this statement.} by studying full nonlinear Hamiltonian for the
relevant conformal and helicity-0 longitudinal modes with no spatial dependence, which is the dimensionally reduced diagonal Ansatz of \cite{IgKa}. We show, however, that within the same Ansatz, depending on the choice of the field parametrization, a perturbative decomposition in terms of the conformal and helicity-0 modes invariably leads
to a) equations of motion with higher-than-second-order time derivatives in agreement with the claim of \cite{RG} or b) a ghost already at the quadratic level. We emphasize that the results of \cite{RG,Ber2,RG1} are obtained in the context of an {\em intrinsically perturbative} field parametrization, which parametrization does not appear to posses a non-perturbative generalization, and all ``no-go" results stemming therefrom appear to be mere artifacts of perturbative expansion.

{}We also revisit the gravitational Higgs mechanism in the simplest case with no higher derivative couplings in the scalar sector,
first discussed in \cite{thooft}, which does {\em not} correspond to the Fierz-Pauli
mass term at the quadratic level. Nonetheless, non-perturbatively, even this case appears to ``resum'' into a theory with
a positive definite Hamiltonian.\footnote{So does a continuous set smoothly connecting this case to the aforesaid Fierz-Pauli case.} We argue that this is the case using our dimensionally reduced diagonal Ansatz. We also reproduce the same result using the full Hamiltonian analysis of
\cite{JK}, thereby validating our Ansatz.

{}The essence of our results is well-illustrated by the following simple ``toy'' example:
\begin{equation}
 {\cal H} = {\mu^2\over \sqrt{p^2 + \mu^2}}~,
\end{equation}
where ${\cal H}$ is the Hamiltonian, $p$ is the canonical conjugate momentum, and $\mu$ is a parameter. Non-perturbatively, this Hamiltonian is positive definite. On the other hand, in the ``weak-field'', small $p^2$ approximation, perturbatively there is a ghost as the kinetic term proportional to $p^2$ has a wrong sign: ${\cal H} = \mu - p^2/2\mu + \dots$; simply put, in the above example the weak-field approximation is invalid in the regime where the fake perturbative ``ghost'', which is merely an artifact of linearization, would destabilize the background. The same appears to be the case in the gravitational Higgs mechanism examples we study in this paper.

{}To summarize, {\em non-perturbatively}, the gravitational Higgs mechanism appears to be free of ghosts.\footnote{However, in this paper we do not attempt to address the question of whether there is any superluminal propagation of signals or the related issue of causality. In this regard, we emphasize that the recent ``no-go" results of \cite{Gruzinov} are obtained in the context of the aforesaid {\em intrinsically perturbative} field parametrization, directly rely on the results of \cite{RG}, and do not apply to the full non-perturbative definition of the gravitational Higgs mechanism. To see if there is any superluminal propagation of signals in the full non-perturbative theory, it appears that one might have to develop some new non-perturbative methods, which is clearly beyond the scope of this paper. The non-Fierz-Pauli model of \cite{thooft} is the ``least non-perturbative" and might provide a fruitful testing ground in this context.}.

{}The rest of the paper is organized as follows. In Sections 2 and 3 we discuss
the gravitational Higgs mechanism in Minkowski background, which results in
massive gravity with the Fierz-Pauli mass term for the appropriately tuned
cosmological constant. In Section 4 we derive the Hamiltonian for
the relevant metric modes and show that it is bounded from below. In Section 5 we revisit the
simplest case without higher derivative couplings in the scalar sector and show that there too the
Hamiltonian is positive definite, in spite of the quadratic truncation not
being of the Fierz-Pauli form.

\section{Minkowski Solutions}

{}Consider the induced metric for the scalar sector:
\begin{equation}
 Y_{MN} = Z_{AB} \nabla_M\phi^A \nabla_N\phi^B~.
\end{equation}
Here $M = 0, \dots, (D-1)$ is a space-time index, and $A = 0,\dots, (D-1)$ is
a global index.
We will choose the scalar metric $Z_{AB}$ to be the Minkowski metric:
\begin{equation}
 Z_{AB} = \eta_{AB}~.
\end{equation}
Let
\begin{equation}\label{Y}
 Y\equiv Y_{MN}G^{MN}~.
\end{equation}
The following action, albeit not the most general\footnote{One can consider a
more general setup where the scalar action is constructed not just from $Y$,
but from $Y_{MN}$, $G_{MN}$ and $\epsilon_{M_0\dots M_{D-1}}$, see, {\em e.g.},
\cite{ZK1,Demir,Cham1,Oda2,IgKa}. However, a simple action containing a scalar
function $V(Y)$ suffices to capture all qualitative features of gravitational
Higgs mechanism. In particular, if this function is quadratic as in
(\ref{quad.pot}), the cosmological constant $\Lambda$ must be negative in the
context of the Minkowski background (but not in the de Sitter case -- see
\cite{IgKa}); however, generically there is no restriction on $\Lambda$, which
can be positive, negative or zero even in the context of the Minkowski
background, once we allow cubic and/or higher order terms in $V(Y)$, or
consider non-polynomial $V(Y)$. As a side remark, let us note that no choice of
polynomial $V(Y)$ constructed from $Y$ only reduces in the decoupling limit to the theories studied in \cite{RG}
({\it cf.} \cite{Ber2}).},
will serve our purpose here:
\begin{equation}
 S_Y = M_P^{D-2}\int d^Dx \sqrt{-G}\left[ R - V(Y)\right]~,
 \label{actionphiY}
\end{equation}
where {\em a priori} the ``potential'' $V(Y)$ is a generic function of $Y$.

{}The equations of motion read:
\begin{eqnarray}
 \label{phiY}
 && \nabla^M\left(V^\prime(Y) \nabla_M \phi^A\right) = 0~,\\
 \label{einsteinY}
 && R_{MN} - {1\over 2}G_{MN} R = V^\prime(Y) Y_{MN}
 -{1\over 2}G_{MN} V(Y)~,
\end{eqnarray}
where prime denotes derivative w.r.t.~$Y$. Multiplying (\ref{phiY}) by
$Z_{AB} \nabla_S\phi^B$ and contracting indices, we
can rewrite the scalar equations of motion as follows:
\begin{equation}\label{phiY.1}
 \partial_M\left[\sqrt{-G} V^\prime(Y) G^{MN}Y_{NS}\right]
- {1\over 2}\sqrt{-G} V^\prime(Y) G^{MN}\partial_S Y_{MN} = 0~.
\end{equation}
Since the theory possesses full diffeomorphism symmetry, (\ref{phiY.1}) and
(\ref{einsteinY}) are not all independent but linearly related due to Bianchi
identities. Thus, multiplying (\ref{einsteinY}) by $\sqrt{-G}$, differentiating
w.r.t.~$\nabla^N$ and contracting indices we arrive at (\ref{phiY.1}).

{}We are interested in finding solutions of the form:
\begin{eqnarray}\label{solphiY}
 &&\phi^A = m~{\delta^A}_M~x^M~,\\
 \label{solGY}
 &&G_{MN} = \eta_{MN}~,
\end{eqnarray}
where $m$ is a mass-scale parameter. The equations of motion (\ref{einsteinY})
imply that
\begin{equation}\label{Ym}
 Y_* \equiv D~m^2
\end{equation}
is the solution of the following equation
\begin{equation}\label{cosm.const}
 V(Y_*) = {2\over D}~Y_* V^\prime(Y_*)~,
\end{equation}
which determines the mass scale $m$.

\section{Massive Gravity}

{}In this section, following \cite{ZK1}, we study linearized fluctuations in
the background given by (\ref{solphiY}) and (\ref{solGY}). Since
diffeomorphisms are broken spontaneously, the equations of motion are invariant
under the full diffeomorphism invariance. The scalar fluctuations $\varphi^A$
can therefore be gauged away using the diffeomorphisms:
\begin{equation}\label{diffphiY}
 \delta\varphi^A =\nabla_M \phi^A \xi^M = m~{\delta^A}_M ~\xi^M~.
\end{equation}
However, once we gauge away the scalars, diffeomorphisms
can no longer be used to gauge away any of the graviton components $h_{MN}$
defined as:
\begin{equation}
 G_{MN} = \eta_{MN} + h_{MN}~.
\end{equation}
Moreover, we will use the notation $h \equiv \eta^{MN} h_{MN}$.

{}After setting $\varphi^A = 0$, we have
\begin{eqnarray}
 && Y_{MN} = m^2 \eta_{MN}~,\\
 && Y = Y_{MN} G^{MN} = m^2 \left[D - h + \dots\right] = Y_* - m^2 h + \dots~,
\end{eqnarray}
where the ellipses stand for higher order terms in $h_{MN}$.

{}Due to diffeomorphism invariance, the scalar equations of motion (\ref{phiY})
are related to (\ref{einsteinY}) via Bianchi identities.
We will therefore focus on (\ref{einsteinY}). Let us first rewrite it as
follows:
\begin{eqnarray}
 &&R_{MN} - {1\over 2}G_{MN} R = \nonumber\\
 &&m^2 \left[ \eta_{MN} V^\prime(Y) - G_{MN} V^\prime(Y_*) \right] -
 {1\over 2}G_{MN}\left[V(Y) - V(Y_*)\right]~.
\end{eqnarray}
Linearizing the r.h.s. of this equation, we obtain:
\begin{eqnarray}
 R_{MN} - {1\over 2}G_{MN} R = {M^2\over 2} \left[\eta_{MN} h -
\zeta h_{MN}\right] +\dots~,
\end{eqnarray}
where
\begin{eqnarray}
 &&M^2 \equiv m^2 V^\prime(Y_*) - 2 m^4 V^{\prime\prime}(Y_*)~,\\
 &&\zeta M^2 \equiv 2 m^2 V^\prime(Y_*)~.
\end{eqnarray}
This corresponds to adding a graviton mass term of the form
\begin{equation}
 -{M^2\over 4} \left[\zeta h_{MN}h^{MN} - h^2\right]
\end{equation}
to the Einstein-Hilbert action,
and the Fierz-Pauli combination corresponds to taking $\zeta = 1$. This occurs
for a special class of potentials with
\begin{equation}\label{tune-V}
 V^\prime(Y_*) = -{2\over D} Y_* V^{\prime\prime}(Y_*)~.
\end{equation}
Thus, as we see, we can obtain the Fierz-Pauli combination of the mass term
for the graviton if we tune {\em one} combination of couplings. In fact, this
tuning is nothing but the tuning of the cosmological constant -- indeed,
(\ref{tune-V}) relates the cosmological constant to higher derivative
couplings.

{}Thus, consider a simple example:
\begin{equation}\label{quad.pot}
 V = \Lambda + Y + \lambda Y^2~.
\end{equation}
The first term is the cosmological constant, the second term is the kinetic
term for the scalars (which can always be normalized such that the
corresponding coefficient is 1 by normalizing the scalars $\phi^A$ accordingly)
, and the third term is a four-derivative term. We then have:
\begin{equation}
 Y_* = -{D\over{2(D+2)}}~\lambda^{-1}~,
\end{equation}
which relates the mass parameter $m$ to the higher derivative coupling
$\lambda$:
\begin{equation}
 m^2 = Y_* / D = -{1\over{2(D+2)}}~ \lambda^{-1}~,
\end{equation}
and the graviton mass is given by:
\begin{equation}
 M^2 = -{2\over{(D+2)^2}}~ \lambda^{-1}~.
\end{equation}
Note that we must have $\lambda < 0$. Moreover, we have:
\begin{equation}\label{LL}
 \Lambda = {{D^2 + 4D - 8}\over {4(D+2)^2}}~\lambda^{-1}~.
\end{equation}
So, the cosmological constant in this case must be negative, which is due to
the choice of the potential (\ref{quad.pot}); however, as we already emphasized
above, for generic choices of the potential there is no restriction on the
cosmological constant, which can be positive, negative or zero.

\section{Is there a Ghost?}\label{sec:gh}

{}The purpose of this section is to argue that the full nonlinear theory of
massive gravity in Minkowski space via gravitational Higgs mechanism is free of
ghosts. We will do this by studying the full nonlinear action for the relevant
modes, which we identify next. In particular, we will argue that no negative
norm state is present for these modes.

{}Let us note that, once we gauge away the scalars, the full nonlinear action
becomes:
\begin{equation}\label{actionphiG}
S_G=M_P^{D-2}\int d^Dx\sqrt{-G}\left[R-{\widetilde V}(\eta^{MN}G_{MN})\right]~,
\end{equation}
where ${\widetilde V}(\zeta) \equiv V(m^2\zeta)$.

{}To identify the relevant modes in the full nonlinear theory, let us note that
in the linearized theory the potentially ``troublesome'' mode is the
longitudinal helicity-0 mode $\rho$. However, we must also include the
conformal mode $\omega$ as there is kinetic mixing between $\rho$ and $\omega$.
In fact, $\rho$ and $\omega$ are not independent but are related via Bianchi
identities. Therefore, in the linearized language one must look at the modes of
the form
\begin{equation}\label{param.lin}
 h_{MN} = \eta_{MN}~\omega + \nabla_M\nabla_N \rho~.
\end{equation}
Furthermore, based on symmetry considerations, namely, the $SO(D-1)$ invariance
in the spatial directions,\footnote{Indeed, negative norm states cannot arise from purely space-like components or spatial derivatives, and are due to time-like components and/or time derivatives.} we can focus on field configurations independent of
spatial coordinates \cite{IgKa}. Indeed, for our purposes here we can compactify the
spatial coordinates on a torus $T^{D-1}$ and disregard the Kaluza-Klein modes.
This way we reduce the $D$-dimensional theory to a classical mechanical system,
which suffices for our purposes here. Indeed, with proper care (see \cite{IgKa}), if
there is a negative norm state in the uncompactified theory, it will be visible
in its compactified version, and vice-versa.

{}Let us therefore consider field configurations of the form:
\begin{equation}\label{param.full}
 G^{MN} = {\rm diag}(g(t)~\eta^{00}, f(t)~\eta^{ii})~,
\end{equation}
where $g(t)$ and $f(t)$ are functions of time $t$ only. The action
(\ref{actionphiG}) then reduces as follows:
\begin{eqnarray}\label{compact}
 S_G = -\kappa \int dt ~g^{-{1\over 2}} f^{-{{D-1}\over 2}}
\left\{\gamma g U^2 + {\widetilde V}(g + \Omega)\right\}~,
\end{eqnarray}
where
\begin{eqnarray}
 &&\kappa\equiv {M_P^{D-2} W_{D-1}}~,\\
 &&\gamma\equiv (D-1)(D-2)~,\\
 &&U\equiv {1\over 2}\partial_t\ln(f)~,\\
 &&\Omega\equiv (D-1)f~,
\end{eqnarray}
and $W_{D-1}$ is the volume in the spatial dimensions ({\em i.e.}, the volume
of $T^{D-1}$). Note that $g$ is a Lagrange multiplier. The goal is to integrate
out $g$ and obtain the corresponding action for $f$. It is then this action
that we should test for the presence of a negative norm state.

{}The equation of motion for $g$ reads:
\begin{equation}\label{geq}
 {\widetilde V}(g+\Omega) - 2g{\widetilde V}^\prime(g +\Omega) = \gamma g U^2~.
\end{equation}
The following discussion can be straightforwardly generalized to general
${\widetilde V}$. However, for our purposes here it will suffice to consider
quadratic ${\widetilde V}$ corresponding to (\ref{quad.pot}). We then have:
\begin{equation}\label{eqn.g}
 3\lambda m^2 g^2 + \left[1 + 2\lambda m^2\Omega +
{\gamma \over{m^2}}~U^2\right] g -
 \left\{{\Lambda\over m^2}+\Omega\left[1+\lambda m^2\Omega\right]\right\}=0~.
\end{equation}
We can therefore express $g$ in terms of $f$ and $\partial_t \ln(f)$:
\begin{eqnarray}
 6\lambda m^2 g = &&-\left[1 + 2\lambda m^2\Omega +
{\gamma \over m^2}~U^2\right] +\nonumber\\
 &&\sqrt{\left[1 + 2\lambda m^2\Omega + {\gamma \over m^2}~U^2\right]^2
+ 12\lambda m^2\left\{{\Lambda\over m^2}
+ \Omega\left[1 + \lambda m^2\Omega\right]\right\}}~,\label{sol.g}
\end{eqnarray}
where the branch is fixed by the requirement that $g \equiv 1$ when
$f \equiv 1$. Substituting the so expressed $g$ into (\ref{compact}), we
obtain an action which is a nonlinear functional of $f$ and
$\partial_t \ln(f)$.

{}For our purposes here it is more convenient to work with the canonical
variable $q$, where
\begin{eqnarray}
 &&q\equiv \ln(f)~,\\
 &&\Omega = (D-1)e^{q}~,\\
 &&U = {1\over 2} \partial_t q~,
\end{eqnarray}
and the action reads:
\begin{eqnarray}\label{compact1}
 S_G = \int dt~L = -\kappa \int {dt} ~g^{-{1\over 2}} e^{-{{D-1}\over 2}q}
\left\{\gamma g U^2 + {\widetilde V}(g + \Omega)\right\}~,
\end{eqnarray}
where $L$ is the Lagrangian. This action corresponds to a classical mechanical
system with a lagrange multiplier $g$.

{}Next, the conjugate momentum is given by
\begin{eqnarray}
 &&p = {{\partial L} \over {\partial(\partial_t q)}} = -\kappa e^{-{{D-1}\over
2}q} \times \nonumber\\
 && \times \left\{{1\over 2} g^{-{1\over 2}} {\hat g} \gamma U^2 - {1\over 2}
g^{-{3\over 2}} {\hat g} {\widetilde V}(g + \Omega) + g^{-{1\over 2}} {\hat g}
{\widetilde V}^\prime(g + \Omega) + g^{{1\over 2}}  \gamma U\right\}~,
\label{momentum}
\end{eqnarray}
where
\begin{equation}
 {\hat g} \equiv {{\partial g} \over {\partial(\partial_\tau q)}}~.
\end{equation}
Using (\ref{geq}), (\ref{momentum}) simplifies to
\begin{equation}
 p = -\kappa e^{-{{D-1}\over 2}q} g^{{1\over 2}} \gamma U~,
\end{equation}
and the Hamiltonian is given by
\begin{equation}
 {\cal H} = p~\partial_t q - L = -\kappa g^{-{1\over 2}} e^{-{{D-1}\over 2}q}
\left[\gamma g U^2 - {\widetilde V}(g + \Omega)\right]~.
\end{equation}
We can now see if this Hamiltonian is bounded from below.

{}First, using (\ref{geq}), we have:
\begin{equation}
 {\cal H} = 2 \kappa g^{{1\over 2}} e^{-{{D-1}\over 2}q}
{\widetilde V}^\prime(g + \Omega) =
 2 m^2 \kappa g^{{1\over 2}} e^{-{{D-1}\over 2}q}\left[1 +
2\lambda m^2(g + \Omega)\right]~.
\end{equation}
Using (\ref{sol.g}), we can rewrite this Hamiltonian as follows:
\begin{equation}
 {\cal H} = {2\over 3} m^2 \kappa g^{{1\over 2}} e^{-{{D-1}\over 2}q}
\left[X - Z\right]~,
\end{equation}
where
\begin{eqnarray}
 &&X \equiv \sqrt{\left[1 + 2\lambda m^2\Omega + {\gamma\over m^2}~U^2\right]^2
+ 12\lambda m^2\left\{{\Lambda\over m^2} +
\Omega\left[1 + \lambda m^2\Omega\right]\right\}}~,\\
 &&Z\equiv {\gamma\over m^2}~U^2 - 4\lambda m^2\Omega - 2~.
\end{eqnarray}
The presence of a ghost would imply that the Hamiltonian is unbounded from
below for large values of $U^2$ (recall that $U^2$ contains the ``kinetic''
term). However, it is not difficult to show that this Hamiltonian suffers from
no such pathology. Indeed, we can rewrite it as follows:
\begin{eqnarray}
 {\cal H} &=& {2\over 3} m^2 \kappa g^{{1\over 2}}
e^{-{{D-1}\over 2}q}~{{X^2 - Z^2}\over {X + Z}} = \nonumber\\
 &&2 m^2 \kappa  g^{{1\over 2}} e^{-{{D-1}\over 2}q}~
{{4\lambda\Lambda - 1 + 2 (\gamma / m^2)~U^2
 \left[1 + 2\lambda m^2 \Omega\right]} \over {X + Z}}~,
\end{eqnarray}
which in the large $U^2$ limit reads:
\begin{equation}
 {\cal H} = 2 m^2 \kappa  g^{{1\over 2}} e^{-{{D-1}\over 2}q}
\left[1 + 2\lambda m^2 \Omega\right] +{\cal O}(1/U^2)~.
\end{equation}
Furthermore, from (\ref{sol.g}) we have
\begin{equation}
 6\lambda m^2 g = {{X^2 - Q^2} \over{X + Q}} = 12 ~{\lambda\Lambda +
\lambda m^2\Omega\left[1 + \lambda m^2\Omega\right] \over{X + Q}} =
{\cal O}(1/U^2)~,
\end{equation}
where
\begin{equation}
 Q\equiv {\gamma\over m^2}~U^2 + 2\lambda m^2\Omega + 1~.
\end{equation}
So, in the large $U^2$ limit the Hamiltonian actually vanishes.

{}Note that the above argument implicitly assumes that $\Omega$ is bounded from
above. This is indeed the case as $g$ must be at least non-negative, which
implies that
\begin{equation}
 \lambda\Lambda + \lambda m^2\Omega\left[1 + \lambda m^2\Omega\right] \leq 0~,
\end{equation}
and $\Omega$ is bounded as follows (note that we must have $\Omega \geq 0$):
\begin{equation}
 \mbox{max}\left(0~,~ -{{1 - \sqrt{1 - 4\lambda\Lambda}}\over 2\lambda m^2}
\right)\leq\Omega\leq -{{1 + \sqrt{1 - 4\lambda\Lambda}}\over 2\lambda m^2}~,
\end{equation}
and we must further have
\begin{equation}\label{LL1}
 \Lambda \geq {1\over 4\lambda}~,
\end{equation}
which is always satisfied due to (\ref{LL}) as $\lambda < 0$.

{}Thus, as we see, there appears to be no ghost in the full nonlinear theory.
The key ingredient here is the parametrization of the conformal and helicity-0
longitudinal modes. We have been working with (\ref{param.full}), while the
results of \cite{RG, Ber2} apply to (\ref{param.lin}). The difference between the
two is that (\ref{param.full}) has no derivatives. Thus, at the quadratic
order, the second derivatives introduced by the parametrization corresponding
to (\ref{param.lin}) can be integrated by parts to arrive at an action
containing only first derivatives of $\omega$ and $\rho$. However, we have
explicitly checked that already at the cubic level the second derivatives
introduced by the parametrization corresponding to (\ref{param.lin}) cannot be
integrated by parts, so the resulting action invariably includes terms with
second derivatives of $\rho$. This is clearly problematic already at the cubic
level and suggests that the parametrization corresponding to (\ref{param.lin})
cannot be used beyond the linearized approximation. Indeed, a nonlinear
completion of (\ref{param.lin}) is given by:
\begin{equation}
 G^{MN} = \eta^{MN} f + \nabla^M\nabla^N u~.
\end{equation}
where the covariant derivative is defined w.r.t. the metric $\eta_{MN}$
(this choice does not affect our discussion here). Note, however, that such a
parametrization of the metric is rather problematic in the context of the full
nonlinear theory (\ref{actionphiG}) as it introduces higher derivative terms
in $u$, which should therefore not be used as the canonical variable in the
full nonlinear theory. This suggests that our parametrization
(\ref{param.full}) is indeed adequate (while our results here suggest that the perturbative field parametrization underlying the results of \cite{Ber2} does not appear to be adequate in the full non-perturbative theory).

\section{The Linear Potential Revisited }

{}In this section we consider the linear potential
\beq\label{linV}
 V(Y) = \Lambda + Y~.
\eeq
This example was discussed in \cite{thooft} and worked out in detail
in the Hamiltonian formalism in \cite{JK}. The potential
(\ref{linV}) does not satisfy the condition (\ref{tune-V}) and, therefore,
does not lead to the Fierz-Pauli action at the quadratic order; consequently, it
propagates a negative norm excitation at that order in perturbation theory.
Nonetheless, we will show that the full non-perturbative Hamiltonian is bounded
from below. This is shown in two ways: In the dimensionally reduced parametrization of
(\ref{param.full}), and in the full Hamiltonian analysis of \cite{JK}.
This supports our conclusion that the Ansatz (\ref{param.full}) is fully adequate for analyzing unitarity.

{}Recalling the definitions of Section \ref{sec:gh}, we have
${\widetilde V}(x) = \Lambda + m^2 x$, and the $g$ equation of motion gives
\beq
g = (\Lambda + m^2 \Omega) / (m^2 + \gamma U^2),
\eeq
thus, the Hamiltonian reads:
\beq\label{linH}
{\cal H} = 2\kappa m^2~e^{-\frac{D-1}{2}q} g^{1/2}~,
\eeq
which is in general positive definite (assuming $g > 0$, which is the case for
$f > -\Lambda / (D-1)m^2 = (D-2)/(D-1)$, where we have used (\ref{Ym}) and (\ref{cosm.const})).
Therefore, as for the quadratic potential of the previous section, it appears
that non-perturbatively there is no ghost.
Furthermore, let us point out that perturbatively there is a ghost, since in a
``weak-field'', small $U^2$ expansion the kinetic term proportional to $U^2$ has
a wrong sign. This is similar to instructive toy examples discussed in Section 1 hereof as well as Appendix A
of \cite{IgKa}.

{}We arrive at the same result in full generality by introducing the lapse
constraint into the gauge-fixed Hamiltonian derived for this potential in
\cite{JK}. The result reads, in their notation,
\beq
{\cal H}_{fix} =
\frac{\sqrt{\det h}}{8\pi G}\sqrt{\frac{1}{N^2} +
\left(\frac{8\pi G}{\sqrt{\det h}}\right)^2
{\cal H}_i^{GR}\delta^{ij}{\cal H}_j^{GR}}~,
\eeq
which is positive definite, and reduces to (\ref{linH}) for the Ansatz
(\ref{param.full}) since, in that case, ${\cal H}_i^{GR}=0$, $1/N^2=g$ and
$\det h=1/f^{D-1}$.

{}In fact, there is nothing ``special'' about the linear potential (\ref{linV}) in terms of unitarity. The discussion in Section 4 is completely independent of (\ref{tune-V}) (with the exception of the reference to (\ref{LL}) immediately following (\ref{LL1}), which, however, does not affect our discussion here -- see below), which is the condition corresponding to having the Fierz-Pauli term at the quadratic order. In particular, the discussion in Section 4 and its conclusion that non-perturbatively there appears to be no negative norm state are valid for all values of $\lambda$, including those smoothly interpolating between (\ref{LL}) (which corresponds to the Fierz-Pauli term) and $\lambda = 0$, which corresponds to the linear potential (\ref{linV}) as long as (\ref{LL1}) is satisfied along the interpolation path (and this condition does not pose an obstruction to such interpolation). Simply put, non-perturbatively there appears to be no negative norm state for a continuous family of models arising out of the gravitational Higgs mechanism, and the linear potential point and the Fierz-Pauli point are not any special in this regard.

{}The fact that non-Fierz-Pauli points appear to be unitary deserves further elaboration. Thus, in the gravitational Higgs mechanism we start with massless gravity with $D(D-3)/2$ degrees of freedom coupled to $D$ scalars, one of which is time-like. As was explained in \cite{ZK1}, at the Fierz-Pauli point (\ref{tune-V}) the time-like scalar does not propagate, so in the scalar sector we have only $(D-1)$ propagating space-like degrees of freedom, which are eaten in the gravitational Higgs mechanism producing $(D + 1)(D - 2)/2 ( = D(D-3)/2 + (D-1))$ degrees of freedom for the massive graviton, all of which are unitary. What about the non-Fierz-Pauli points? Here we have an extra time-like scalar degree of freedom, which in the perturbative language in the Higgs phase results in a ghost-like propagating degree of freedom, namely, the trace of the graviton $h = {h^M}_M$. However, {\em non-perturbatively} the Hamiltonian is positive-definite and there is no ghost, while if the Hamiltonian is expanded perturbatively, as we saw above, a ghost invariably appears. Simply put, the weak-field approximation is invalid in the regime where the fake perturbative ``ghost'', which is merely an artifact of linearization, would destabilize the background.\footnote{In this regard, let us note a difference between the gravitational Higgs mechanism and its gauge theory counterpart. In the latter scalar vacuum expectation values are constant, while in the former they depend linearly on space-time coordinates: (\ref{solphiY}). In fact, the background is not even static. It would take infinite energy to destabilize such a background. This is reminiscent to infinite-tension domain walls discussed in \cite{DS, ZK3}.}

\section*{Acknowledgements}
The work of AI is supported by a David and Lucile Packard Foundation Fellowship
for Science and Engineering and NSF grant PHY-0758032.


\end{document}